\documentclass[12pt]{article}

\usepackage[left=2.8cm,right=2.8cm,top=2.8cm,bottom=2.8cm]{geometry}

\usepackage[colorlinks=true,linkcolor=black,citecolor=black,urlcolor=blue,filecolor=black]{hyperref}

 \csname
@addtoreset\endcsname{equation}{section}

\usepackage[utf8]{inputenc}
\usepackage{amsmath}
\usepackage{amsfonts}
\usepackage{amssymb}
\usepackage{stmaryrd}
\usepackage{mathtools}

\usepackage[sort]{cite}
\usepackage[usenames,table]{xcolor}
\usepackage{dsfont}
\usepackage{mathrsfs}
\usepackage{comment}
\usepackage{verbatim}

\def\cE{{\cal E}}

\def\cL{{\cal L}}

\def\cR{{\cal R}}

\def\a{\alpha} 
  
\def\g{\gamma} 
\def\G{\Gamma}
\def\d{\delta} 

\def\e{\epsilon} 
\def\ve{\varepsilon}

\def\h{\eta}

\def\La{\Lambda}
\def\m{\mu}
\def\n{\nu}
\def\p{\pi}

\def\r{\rho}
\def\s{\sigma}

\def\t{\tau}

\def\O{\Omega}
\newcommand{\be}{\begin{equation}}
\newcommand{\ee}{\end{equation}}
\newcommand{\bal}{\begin{aligned}}
\newcommand{\eal}{\end{aligned}}
\def\pr{\partial}

\begin{document}

\vspace{30pt}

\begin{center}

%%%%%%%%%%%%%%%%%%%%%%%%%%%%%%%%%%%%%%%%%%%%%%%%%%%%%%%%

{\Large\sc Cartan-like formulation \\[10pt] of electric Carrollian gravity}

\vspace{-5pt}
\par\noindent\rule{350pt}{0.4pt}

%%%%%%%%%%%%%%%%%%%%%%%%%%%%%%%%%%%%%%%%%%%%%%%%%%%%%%%%

\vspace{20pt}
{\sc 
Simon Pekar${}^{\; a,}$\footnote{Centre National de la Recherche Scientifique, Unit\'e Mixte de Recherche UMR 7644.},
Alfredo~P\'erez${}^{\; b, c}$ and
Patricio~Salgado-Rebolledo${}^{\; d, e}$
}

\vspace{8pt}

${}^a${\it\small 
Centre de Physique Th\'eorique - CPHT\\ \'Ecole polytechnique, CNRS\\ Institut Polytechnique de Paris, 91120 Palaiseau Cedex, France}
\vspace{4pt}

${}^b${\it\small
Centro de Estudios Cient\'ificos (CECs), Avenida Arturo Prat 514, Valdivia, Chile}
\vspace{4pt}

${}^c${\it\small
Facultad de Ingenier\'ia, Arquitectura y Dise\~{n}o, Universidad San Sebasti\'an, sede Valdivia, General Lagos 1163, Valdivia 5110693, Chile}
\vspace{4pt}

${}^d${\it\small
Institute for Theoretical Physics, TU Wien, Wiedner Hauptstr.~8-10/136, \\
A-1040 Vienna, Austria}
\vspace{5pt}

${}^e${\it\small
Institute of Theoretical Physics, Wroc\l aw University of Science and Technology,\\
50-370 Wroc\l aw, Poland}
\vspace{5pt}

{\tt\small 
\href{mailto:simon.pekar@polytechnique.edu}{simon.pekar@polytechnique.edu},
\href{mailto:alfredo.perez@uss.cl}{alfredo.perez@uss.cl},
\href{mailto:patricio.salgado-rebolledo@pwr.edu.pl}{psalgadoreb@hep.itp.tuwien.ac.at}
}

%%%%%%%%%%%%%%%%%%%%%%%%%%%%%%%%%%%%%%%%%%%%%%%%%%%%%%%%
\vspace{30pt} {\sc\large Abstract}  \end{center}

\noindent

We present a Cartan-like first-order action principle for electric Carrollian gravity. The action is invariant under the local homogeneous Carroll group, albeit in a different representation than the one obtained by gauging the Carroll algebra. Additionally, we show that this first-order action can be derived from a smooth Carrollian limit of the Einstein-Cartan action. The connection with the Hamiltonian and metric forms of the action for electric Carrollian gravity, as well as with previous works in the literature, is also discussed.

\newpage

%%%%%%%%%%%%%%%%%%%%%%%%%%%%%%%%%%%%%%%%%%%%%%%%%%%%%%%%
%%%%%%%%%%%%%%%%%%%%%%%% TEXT %%%%%%%%%%%%%%%%%%%%%%%%%%
%%%%%%%%%%%%%%%%%%%%%%%%%%%%%%%%%%%%%%%%%%%%%%%%%%%%%%%%

\tableofcontents

%%%%%%%%%%%%%%%%%%%%%%%%%%%%%%%%%%%%%%%%%%%%%%%%%%%%%%%%%%%%%%
\section{Introduction} \label{sec:introduction}
%%%%%%%%%%%%%%%%%%%%%%%%%%%%%%%%%%%%%%%%%%%%%%%%%%%%%%%%%%%%%%

The Carroll algebra is defined through an In\"on\"u-Wigner contraction
of the Poincar\'e algebra in the limit when the speed of light approaches
zero ($c\rightarrow0$) \cite{Levy:1965, SenGupta:1966}. In recent years, there has been an increasing interest in Carrollian physics mainly due to the isomorphism between the conformal extension of the Carroll symmetry \cite{Duval:2014uva} and the asymptotic symmetry algebra of asymptotically flat spacetimes \cite{Bondi:1962,Sachs:1962,Barnich:2009se}. This is a consequence of the Carrollian structure that naturally emerges at null infinity of asymptotically flat spacetimes, which was later understood to be a generic property of null hypersurfaces \cite{Hartong:2015usd,Ciambelli:2018ojf}. Physical systems exhibiting Carrollian symmetries have found applications
in various fields, such as black holes physics \cite{Penna:2018gfx,Donnay:2019jiz,Gray:2022svz,Redondo-Yuste:2022czg,Bicak:2023rsz,Ecker:2023uwm,Bagchi:2023cfp,Aggarwal:2024gfb}, flat space holography \cite{Donnay:2022aba,Donnay:2022wvx,Bagchi:2022emh,Campoleoni:2023fug,Saha:2023hsl,Salzer:2023jqv,Nguyen:2023vfz,Mason:2023mti,Bekaert:2024itn}, tensionless string theory \cite{Bagchi:2015nca,Bagchi:2022iqb,Fursaev:2023lxq}, the fluid/gravity correspondence \cite{deBoer:2017ing,Ciambelli:2018xat,Freidel:2022bai,Petkou:2022bmz,Armas:2023dcz}, cosmology \cite{deBoer:2021jej,Bagchi:2021qfe}, Hall effects \cite{Marsot:2022imf}, fractons \cite{Casalbuoni:2021fel,Bidussi:2021nmp,Figueroa-OFarrill:2023vbj,Figueroa-OFarrill:2023qty,Pena-Benitez:2023aat,Perez:2023uwt}, and flat bands in condensed matter systems \cite{Bagchi:2022eui}. Furthermore, Carroll physics has recently developed in several directions, including Carrollian gravity \cite{Dautcourt:1997hb, Hartong:2015xda,Bergshoeff:2017btm,Matulich:2019cdo,Ravera:2019ize,Gomis:2019nih,Grumiller:2020elf,Gomis:2020wxp,Perez:2021abf,Hansen:2021fxi,Concha:2021jnn,Lovrekovic:2021dvi,Figueroa-OFarrill:2022mcy,Campoleoni:2022ebj,Miskovic:2023zfz,Sengupta:2022rbd,Musaeus:2023oyp,Guerrieri:2021cdz}, Carroll particles \cite{Bergshoeff:2014jla,Casalbuoni:2023bbh} and Carrollian field theories \cite{Rivera-Betancour:2022lkc,Kasikci:2023zdn,deBoer:2023fnj,Parekh:2023xms,Bergshoeff:2023vfd,Ciambelli:2023xqk,Ecker:2024czx}.

As shown in Ref.~\cite{Henneaux:2021yzg}, relativistic
bosonic theories admit at least two inequivalent Carrollian limits
termed electric and magnetic, respectively. Electric-type field theories
have a long history. Klauder originally studied the electric Carrollian
scalar field theory in the 1960s, referring to it as an ultralocal
scalar field~\cite{Klauder:1970cs}. In gravitation, the electric limit was introduced by Isham~\cite{Isham:1975ur}
within the context of strong gravity (see also \cite{Anderson:2002zn}), and  by Teitelboim
as a zero signature limit of General Relativity~\cite{Teitelboim:1978wv} (see also \cite{Henneaux:1979vn, Henneaux:1981su}). This particular gravitational
theory is also ultralocal, meaning that different points in space
are completely decoupled from each other. This feature makes the theory
useful in describing the dynamics of the gravitational field near
space-like singularities using the Belinsky-Khalatnikov-Lifshitz (BKL)
approach~\cite{Belinsky:1970ew,Henneaux:1982qpq,Belinsky:1982pk,Damour:2002et}. The asymptotic symmetries
of this theory were recently analyzed in Refs.~\cite{Perez:2021abf,Perez:2022jpr,Fuentealba:2022gdx}.

The Hamiltonian action of electric Carrollian gravity takes the form 
\begin{equation}
  I_E[g_{ij}, N_i, N, \pi^{ij}] =\int dt\,d^D x \left[ \pi^{ij}\,\dot g_{ij} - N\,\mathcal H - N^i\,\mathcal H_i \right] , \label{eq:HamElec}
\end{equation}
where
\begin{equation}
    \mathcal H= \frac{16\pi G_E}{\sqrt g}\left(\pi^{ij}\,\pi_{ij}-\frac{1}{D-1}\pi^2\right)+\frac{\sqrt g \Lambda_E}{8\pi G_E} \,,\qquad \mathcal H_i = - 2\,\nabla_j\,\pi_i{}^{j} \,.
\end{equation}
Here $g_{ij}$ is the spatial metric and $\pi^{ij}$ is its conjugate canonical momentum, while $\mathcal H$ and $\mathcal H_i$ represent the Hamiltonian and momentum constraints, respectively. The corresponding Lagrange multipliers, lapse and shift, are denoted by $N$ and $N^{i}$. Here $G_{E}$ and $\Lambda_E$ represent the rescaled Newton's and cosmological
constants. In particular, note that the Hamiltonian constraint $\mathcal H$ does not contain spatial derivatives, reflecting the ultralocal nature of the theory.

A covariant formulation of electric Carrollian gravity in the second
order formalism was provided by Henneaux in Ref.~\cite{Henneaux:1979vn}.
The action principle that describes its dynamics in a $(D+1)-$dimensional spacetime is the following
\begin{equation}
I_{E}=\frac{1}{16\pi G_{E}}\int dt\,d^{D}x\,\mathcal{E}\left[K_{\mu\nu}K^{\mu\nu}-K^{2}-2\Lambda_E\right].\label{eq:IE}
\end{equation}
The volume form is given by $\mathcal{E}\,dt\,d^{D}x$,
while $K_{\mu\nu}$ represents the extrinsic curvature of the spatial
hypersurfaces and $K = g^{\mu\nu} K_{\mu\nu}$ denotes its trace. 

So far, a Cartan-like formulation of electric Carrollian gravity has
remained elusive, mainly because the terms at hand for defining
an action principle through the gauging of the Carroll algebra do
not reproduce the second-order action of the electric theory~\cite{Figueroa-OFarrill:2022mcy}.
This stands in sharp contrast to magnetic Carrollian gravity, where its
description in terms of the gauging of the Carroll algebra is well
understood~\cite{Bergshoeff:2017btm,Campoleoni:2022ebj}.\footnote{Note, however, the proposal of \cite{Hartong:2015xda} in $2+1$ dimensions, where a gauging of the Carroll algebra was shown to lead to a theory containing electric terms of the form of \eqref{eq:IE}.}

One of the main goals of our article is to propose a first-order formulation
of electric Carrollian gravity \`a la Cartan. This first-order action is invariant
under a different representation of the local homogeneous Carroll
transformations than the one obtained by gauging the Carroll algebra.
Therefore, the action is of first-order but is not derived from the
gauging method. This is fully consistent with the previous no-go results~\cite{Figueroa-OFarrill:2022mcy}. 

From a purely geometric perspective, the incompatibility with the gauging method can be understood as follows: as demonstrated in Ref.~\cite{Henneaux:1979vn}, in Carrollian theories where the extrinsic curvature tensor $K_{\mu\nu}$ does not vanish, such as electric Carrollian gravity, only a very limited concept of metric parallel transport exists. It is restricted to purely transverse and longitudinal tensors along the normal direction. On the other hand, the second-order electric Carrollian action \eqref{eq:IE} depends exclusively on well-defined intrinsic geometric quantities. In particular, since there are no covariant derivatives, the action is independent of any kind of (non-metric) connection. From the point of view of the first-order action, we will see that this property is linked to the absence of the spin connection associated with boost generators, as can be seen from Eq.~\eqref{eq:deltav}.

The Cartan-like action for electric Carrollian gravity proposed in this
article can also be derived from a Carrollian limit of the Einstein-Cartan
action. While the Carroll limit remains smooth at the level of the
action principle, it is shown to be discontinuous at the level of
the gauging algebra. This naturally explains why a direct application
of the gauging method to the Carroll algebra does not provide a first-order
formulation of the electric theory.

The plan of the paper is the following: in section~\ref{sec:first-order}, we propose a Cartan-like first-order action in terms of the fields $\left\{e_\mu{}^a,\, \tau_\mu,\, \omega_\mu{}^{ab},\, v_\mu{}^a\right\}$ that are reminiscent of the first-order formulation of magnetic Carrollian gravity obtained from gauging the Carroll algebra. This first-order action is invariant under local transformations that close on the homogeneous Carroll algebra, and we prove its equivalence to the electric action in second-order form. In section~\ref{sec:Electric_action_from_a_limit}, we show how to obtain the previous action from a limit of Einstein-Cartan. 
Finally, we conclude by giving an outlook on our results in section~\ref{sec:conclusions}.

%%%%%%%%%%%%%%%%%%%%%%%%%%%%%%%%%%%%%%%%%%%%%%%%%%%%%%%%%%%%%%
\section{Cartan-like first-order formulation of electric Carrollian gravity} \label{sec:first-order}
%%%%%%%%%%%%%%%%%%%%%%%%%%%%%%%%%%%%%%%%%%%%%%%%%%%%%%%%%%%%%%

%%%%%%%%%%%%%%%%%%%%%%%%%%%%%%%%%%%%%%%%%%%
\subsection{Geometric preliminaries}

The metric properties of Carrollian geometries can be described by a clock form $\tau_{\mu}\,$ and a spatial vielbein $e_\m{}^a\,$. Here, Greek letters denote spacetime indices, $\mu,\nu=0,1,\dots, D$, while lowercase letters $a,b=1,\dots,D$ denote spatial indices on the tangent space. The set $\{\t_\m, e_\m{}^a\}$ is assumed to constitute a basis of the cotangent space. In addition one can introduce the dual vectors $\{n^\m,e^\m{}_a\}$ satisfying
\be
e_\m{}^a\,e^\m{}_b = \d^a{}_b \,,\quad \t_\m\,n^\m = 1 \,,\quad n^\m\,e_\m{}^a = 0 \,,\quad \t_\m\,e^\m{}_a = 0 \,,
\ee
and
\be \label{eq:inverse-vielbein}
e_{\mu}{}^{a}e^{\nu}{}_{a} + \tau_{\mu} n^{\nu}=\delta_{\mu}{}^{\nu} \,,
\ee
i.e. $e_\m{}^a$, $e^\m{}_a$, $\t_\m$ and $n^\m$ provide a resolution of the identity. 

Using these objects, it is possible to construct the following determinant
\be
\mathcal E:=\det\left(\tau_{\mu},e_{\mu}{}^{a}\right) = \frac{1}{D!}\, \epsilon_{a_1\cdots a_D}\epsilon^{\mu_0\cdots \mu_D}\tau_{\mu_0} e_{\mu_1}{}^{a_1}\cdots e_{\mu_D}{}^{a_D} \not= 0 \,.
\ee
The Carrollian metric is then defined as follows
\be
g_{\mu\nu}=\delta_{ab}\,e_\m{}^a\,e_\n{}^b,  
\ee
and satisfies the following condition
\begin{equation}
    g_{\mu\nu}n^{\nu}=0 \,,
\end{equation}
which explicitly shows its degenerate nature. 

The spacetime indices $\m$ and $\n$ are raised and lowered thanks to the transverse co-metric defined by
\be
g^{\m\n} = \delta^{ab}\,e^\m{}_a\,e^\n{}_b \,,
\ee
which is also degenerate (along $\tau$) and is often called the pseudo-inverse of $g_{\m\n}\,$. It satisfies the following condition
\begin{equation}
    g^{\mu\sigma}g_{\sigma\nu}=\delta^{\mu}_{\nu}-n^{\mu}\tau_{\nu} \,.
\end{equation}
Additionally, one can define the extrinsic curvature tensor
\be
K_{\m\n} = -\frac12\,\cL_n g_{\m\n} \,,
\ee
which is transverse and exclusively expressed in terms of metric quantities.

As discussed in the introduction, the concept of parallel transport is limited when the extrinsic curvature does not vanish in a Carrollian manifold \cite{Henneaux:1979vn}, which is precisely the situation in electric Carrollian gravity. To describe its dynamics using the first-order formalism, we will introduce the spin connection associated to rotations, denoted by $\omega_\mu{}^{ab}\,$, and an additional field $v_\mu{}^{a}\,$. It is important to note that the field $v_\mu{}^a$ \textit{is not} the spin connection associated to boost generators (although we will see that it can be seen as originating from its relativistic parent, in the $c \to 0$ procedure detailed in section~\ref{sec:Electric_action_from_a_limit}). In the transition from first to second-order formulations, this field will be completely determined by metric quantities and will be proportional to the extrinsic curvature. In the first-order formulation of electric Carrollian gravity presented in this article, there is no spin connection related to boosts, which is in line with the restrictive notion of parallel transport in this theory.

In summary, the Cartan-like first-order theory is described by the clock form $\tau_{\mu}$, the spatial vielbein $e_\m{}^a$, the spin connection associated with rotations $\omega_\m{}^{ab}$, and an additional one-form $v_\m{}^a$.

%%%%%%%%%%%%%%%%%%%%%%%%%%%%%%%%%%%%%%%%%%%%%%%%%%%%%%%%%%%%%%
\subsection{New representation of the local homogeneous Carroll group and action}

In this section, we present a first-order formulation for electric Carrollian gravity \`a la Cartan. The approach presented here differs significantly from the description of magnetic Carrollian gravity as performed in Refs. \cite{Bergshoeff:2017btm, Campoleoni:2022ebj}, which relies on the gauging of the Carroll algebra.

A crucial demand is that the first-order action remains invariant under local transformations that define a representation of the homogeneous Carroll group, which is generated by spatial rotations and Carroll boosts, and whose Lie algebra is isomorphic to $\mathfrak{iso}(D)\,$. 

The transformation laws of the fields under the local homogeneous Carroll group are postulated to be as follows:
\begin{subequations}
\label{eq:transf_fields}
\begin{align}
\d \t_\m &= - e_\m{}^a\,\r_a \,,\\
\d e_\m{}^a &= - e_\m{}^b\,\lambda^a{}_b \,,\\
\d\omega_\m{}^{ab} &= \pr_\m\lambda^{ab} + 2\,\omega_\m{}^{c[a}\,\lambda^{b]}{}_c + 2\,v_\m{}^{[a}\,\r^{b]} \,,\label{eq:deltaomega}\\
\d v_\m{}^a &= - v_\m{}^b\,\lambda^a{}_b \,,\label{eq:deltav}
\end{align}
\end{subequations}
where here and in the following, square (resp. round) brackets denote antisymmetrisation (resp. symmetrisation) with weight one, i.e. $\lambda_{[ab]} = \frac12 (\lambda_{ab} - \lambda_{ba})\,$. The parameter $\lambda_{[ab]}$ is associated with local rotations, while $\rho^{a}$ is the parameter of local Carroll boosts for the Carrollian geometry $(\tau_\mu,e_\mu{}^a)$.

Note that, although the vielbeine $e_\m{}^a$ and $\t_\m$ transform exactly like as gauge fields that are obtained from the gauging of the Carroll algebra \cite{Hartong:2015xda,Bergshoeff:2017btm,Campoleoni:2022ebj}, the fields $\omega_\m{}^{ab}$ and $v_\m{}^a$ do not. To be more precise, the gauging procedure yields a transformation where the final term on the right-hand side of Eq.~\eqref{eq:deltaomega} is absent. Furthermore, Eq.~\eqref{eq:deltav} shows that the field $v_\m{}^a$ does not transform as a connection, due to the absence of a differential piece proportional to the boost parameter.

The key point is that the commutator of two gauge transformations $\left[\delta_1,\delta_2\right]=\delta_3$ perfectly closes within the homogeneous Carroll algebra according to
\begin{subequations} \label{eq:gauge-algebra}
\begin{align}
\lambda_{\left(3\right)}^{ab} & =\lambda_{\left(1\right)c}^{a}\,\lambda_{\left(2\right)}^{cb} - \lambda_{\left(2\right)c}^{a}\,\lambda_{\left(1\right)}^{cb} \,,\\
\rho_{\left(3\right)}^{a} & =\lambda_{\left(1\right)c}^{a}\,\rho_{\left(2\right)}^{c} - \lambda_{\left(2\right)c}^{a}\,\rho_{\left(1\right)}^{c} \,.
\end{align}
\end{subequations}
Hence, the transformation laws of Eqs.~\eqref{eq:transf_fields} define a representation of this algebra, albeit not being the one obtained by the gauging method.

From the completeness relation of Eq.~\eqref{eq:inverse-vielbein}, one derives the transformation laws of the basis vectors
\begin{subequations}
\label{eq:transf_inv_fields}
\begin{align}
\d e^\m{}_a &= e^\m{}_b\,\lambda^b{}_a + n^\m\,\r_a \,,\\
\d n^\m &= 0 \,.
\end{align}
\end{subequations}
It is evident from the transformations given by Eqs.~\eqref{eq:transf_fields} and \eqref{eq:transf_inv_fields} that the significant geometric entities in the second-order formalism, namely the degenerate metric $g_{\m\n}$ and the vector $n^{\mu}\,$, remain invariant under local homogeneous Carroll transformations, as is also the case for the magnetic theory of Carrollian gravity obtained from gauging the Carroll algebra.

The proposed first-order action of electric Carrollian gravity that is invariant under the local homogeneous Carroll transformations in Eqs.~\eqref{eq:transf_fields} and \eqref{eq:transf_inv_fields} reads
\be \label{eq:electric}
\begin{split}
I_E&[e_\m{}^a, \t_\m, \omega_\m{}^{ab}, v_\m{}^a] \\
&:= \frac{1}{8\pi G_E}\int dt\,d^D x\ \cE \left[ e^\m{}_a\,e^\n{}_b\,v_{[\m}{}^a\,v_{\n]}{}^b - 2\,e^\m{}_a\,n^\n\left(\pr_{[\m}\,v_{\n]}{}^a + \omega_{[\m}{}^{ab}\,v_{\n]}{}_b\right) - \Lambda_E \right] ,
\end{split}
\ee
where $G_E$ is Newton's constant and $\Lambda_E$ plays the role of a Carrollian cosmological constant. In section~\ref{sec:Electric_action_from_a_limit}, we will show that this action can be derived from taking a Carrollian limit of Einstein gravity in the Cartan formalism.

%%%%%%%%%%%%%%%%%%%%%%%%%%%%%%%%%%%%%%%%%%%
\subsection{Equivalence to the electric theory in second-order formulation}

In this subsection we show the equivalence between the first-order action described in Eq.~\eqref{eq:electric} and the second-order action of electric Carrollian gravity.

Firstly, let us remark that not all the components of $\omega_\m{}^{ab}$ enter in the action~\eqref{eq:electric}. Indeed, using the resolution of the identity provided in Eq.~\eqref{eq:inverse-vielbein}, we can write the most general Ansatz for $v_\m{}^a$ and $\omega_\m{}^{ab}$
\begin{subequations}
\begin{align}\label{formofv}
v_\m{}^a &= e_\m{}^b S^a{}_b + \t_\m U^a \,,\\
\omega_\m{}^{ab} &= e_\m{}^c M^{ab}{}_c + \t_\m N^{ab} \,,
\end{align}
\end{subequations}
where $M^{ab}{}_c$ and $N^{ab}$ are zero-forms antisymmetric in $a$ and $b$.
Then, plugging back this decomposition in the action given in Eq.~\eqref{eq:electric}, focusing on the term proportional to $\omega_\m{}^{ab}\,$, we get
\be \label{eq:projection}
- 2\,e^\m{}_a\,n^\n\,\omega_{[\m}{}^{ab}\,v_{\n]}{}_b = - e^\m{}_a\,n^\n\,\omega_\m{}^{ab}\,v_\n{}_b + e^\m{}_a\,n^\n\,\omega_\n{}^{ab}\,v_\m{}_b = - M^{ab}{}_a\,U_b + N^{ab}\,S_{[ba]} \,,
\ee
therefore, only the trace of $M^{ab}{}_c\,$, i.e. $M^{ab}{}_a$ is present in the first-order action. On the other hand, all components of $v_\m{}^a$ are represented in Eq.~\eqref{eq:electric}, since the symmetric part of $S_{ab}$ is also present in the first term $e^\m{}_a\,e^\n{}_b\,v_{[\m}{}^a\,v_{\n]}{}^b = S^a{}_{[b}\,S^b{}_{a]}\,$.

Taking the variation of the action of Eq.~\eqref{eq:electric} with respect to the components of $\omega_\m{}^{ab}$ present in Eq.~\eqref{eq:projection}, which are all Lagrange multipliers, gives rise to the constraints
\be \label{eq:torsion-1}
S^{[ab]} \approx 0 \,,\quad U^a \approx 0 \,,
\ee
which means that $v_\m{}^a$ is only carrying a symmetric piece $S^{(ab)}$ along $e_\m{}^a\,$. This symmetric tensor $S^{(ab)}$ will become the extrinsic curvature upon imposing the next torsion constraint.

Next, we vary the action with respect to all components of $v_\m{}^a\,$. After integrating by parts, getting rid of a total derivative term and some algebraic manipulation, we obtain
\begin{subequations} \label{eq:torsion-2}
\begin{align}
n^\m \left(\partial_{[\m}\,e_{\n]}{}^a + \omega_{[\m}{}^{ab}\,e_{\n]}{}_b + v_{[\m}{}^a\,\t_{\n]} \right) &\approx 0 \,,\\
e^\m{}_a \left(\partial_{[\m}\,e_{\n]}{}^a + \omega_{[\m}{}^{ab}\,e_{\n]}{}_b + v_{[\m}{}^a\,\t_{\n]} \right) &\approx 0 \,.
\end{align}
\end{subequations}
In these two equations of motion, the first two terms within brackets form the usual $\omega_\m{}^{ab}\,-\,$covariant derivative in Carrollian gravity, see \cite{Hartong:2015xda}, and these constraints can be interpreted as projections of a torsionless condition in Carrollian gravity, where the last term of the left-hand side acts as a source of torsion. Upon projecting the previous equations on $e^\n{}_c$ and using the fact that, on-shell, the only non-zero component of $v_\m{}^a$ is $S^{(ab)}\,$, we obtain the following equations
\begin{subequations}
\begin{align}
e^\m{}_a\,\omega_\m{}^{ab} = M^{ab}{}_a &\approx 2\,e^\m{}^b\,e^\n{}_a\,\pr_{[\m}\,e_{\n]}{}^a \,,\\
n^\m\,\omega_\m{}^{ab} = N^{ab} &\approx 2\,n^\m\,e^\n{}^{[a}\,\pr_{[\m}\,e_{\n]}{}^{b]} \,,\\
e^\m{}^{(a} v_\m{}^{b)} = S^{(ab)} &\approx 2\,n^\m\,e^\n{}^{(a}\,\pr_{[\m}\,e_{\n]}{}^{b)} \,.
\end{align}
\end{subequations}
Because of the projection, there is no expression relating the full spin connection $\omega_\m{}^{ab}$ in terms of derivatives of the vielbein, but there is one for $v_\m{}^a$ which takes the form
\be \label{vandomega}
v_\m{}^a \approx - e_\m{}_b\,K^{ab} \,,
\ee
where $K_{ab} = e^\m{}_a\,e^\n{}_b\,K_{\m\n}\,$. We can now replace the previous expressions back into the action and obtain
\be
\label{eq:El_action}
\begin{split}
I_E[g_{\m\n},n^\m] &\approx \frac{1}{8\pi G_E}\int dt\,d^D x\ \cE \left[ -e^\m{}_a\,e^\n{}_b\,v_{[\m}{}^a\,v_{\n]}{}^b - \Lambda_E \right] \\
&\approx \frac{1}{16\pi G_E}\int dt\,d^D x\ \cE \left[ K_{\m\n}\,K^{\m\n} - K^2 - 2\,\Lambda_E \right] ,
\end{split}
\ee
which precisely corresponds to the electric action in the second-order formalism given in Eq.~\eqref{eq:IE}.

As was the case in the magnetic theory \cite{Campoleoni:2022ebj}, we have the possibility starting from Eq.~\eqref{eq:electric} to partially solve the torsion constraints given in Eqs.~\eqref{eq:torsion-1} and \eqref{eq:torsion-2} by performing a variation with respect to $\omega_\m{}^{ab}$ and $v_\m{}^a$ leaving out the symmetric component $S_{\m\n} = v_{(\m}{}^a\,e_{\n)}{}_a\,$ (see Appendix \ref{AppFirstOrder}). This leads to the following alternative covariant first-order form of the electric action 
\be 
I_E[g_{\m\n}, n^\m, P^{\m\n}] \approx \frac{1}{8 \pi G_E} \int dt\,d^D x\ \cE \left[ K_{\m\n} \, P^{\m\n} - G_{\m\n\r\s}\,P^{\m\n}\,P^{\r\s} - \Lambda_E \right] ,
\ee
which was introduced in \cite{Campoleoni:2022ebj}.  Moreover, the electric action can be directly recovered in Hamiltonian form \eqref{eq:HamElec} by expressing $\{\tau_\mu,e_\mu{}^a\}$ and $\{n^\nu,e^\mu{}_a\}$ in terms of ADM variables, which in the time gauge leads to \cite{Henneaux:1979vn}
\be
\bal
&\tau_\mu=(N,0)\,,\quad n^\mu=(N^{-1},-N^i N^{-1}) \,,
\\[5pt]
& e_\mu{}^a =(N^j e_j{}^a, e_i{}^a)\,,\quad e^\mu{}_a =(0, e^i{}_a) \,.
\eal
\ee
and replacing these expressions in the original action \eqref{eq:electric} (see Appendix \ref{AppHamiltonian}).

Note that in Ref. \cite{Sengupta:2022rbd}, a first action for electric Carrollian gravity in Hamiltonian form was also proposed. This action can be seen as an intermediate step between our approach and the final second-order action, where the time gauge was imposed and the `torsion constraint' associated with $v_\mu{}^a$ was solved.

%%%%%%%%%%%%%%%%%%%%%%%%%%%%%%%%%%%%%%%%%%%
\section{Cartan-like electric action from a limit of the Einstein-Cartan action}
\label{sec:Electric_action_from_a_limit}
%%%%%%%%%%%%%%%%%%%%%%%%%%%%%%%%%%%%%%%%%%%

In this section, we derive the first order action for electric Carrollian gravity, as described by Eq.~\eqref{eq:electric}, from an ultra-relativistic limit of the Einstein-Cartan action. This limit differs from that used in Refs. \cite{Bergshoeff:2017btm, Campoleoni:2022ebj}, which lead to magnetic Carrollian gravity. Moreover, it possesses a crucial property: while being well-defined at the level of the action, it is not the case of the gauge algebra. This property of the limit elucidates the challenges encountered when attempting to describe electric Carrollian gravity through the gauging of the Carroll algebra \cite{Figueroa-OFarrill:2022mcy}.  

The first order action of General Relativity is given by
\be \label{Einstein-Cartan}
I[E_\m{}^A, \Omega_\m{}^{AB}] = \frac{c^3}{16 \p G_N} \, \int dt \,d^D x\ E \left[ E^\m{}_A\,E^\n{}_B\, \cR_{\m\n}{}^{AB} - 2\,\La \right] .
\ee
Here, $E_\m{}^A$ denotes the coframe, $\Omega_\m{}^{AB}$ the spin connection and $E=\text{det}\left(E_\m{}^A \right)$, with $A,B=0,1,\dots,D\,$.

The curvature two-form reads
\be \label{Lorentz-curvature}
\cR_{\m\n}{}^{AB} = 2\,\pr_{[\m}\,\O_{\n]}{}^{AB} + 2\,\O_{[\m}{}^{AC}\,\O_{\n]}{}^{DB}\,\h_{CD} \, ,
\ee
where $\h_{CD} = \text{diag}(-1,+1,\ldots,+1)\,$. Newton's constant is denoted by $G_N\,$, and the speed of light by $c\,$. It is convenient to express the speed of light as $c = \hat c\,\e\,$, where $\e$ is a dimensionless parameter. To simplify the notation, we will also set $\hat c = 1\,$, the ultrarelativistic limit then corresponds to taking $\e\rightarrow 0\,$. 

To make contact with the notations of the previous sections, let us consider the following decomposition of the coframe and the spin connection
\be \label{rescalings}
E_\m{}^A
= \left( \e\,\t_\m, e_\m{}^a\right), \qquad
\O_\m{}^{AB} 
= \left( \e\,\omega_\m{}^a, \omega_\m{}^{ab} \right) ,
\ee
where now
\be
E=\epsilon\,\mathcal E. 
\ee
Therefore, the Einstein-Cartan action given in Eq.~\eqref{Einstein-Cartan} can be rewritten as 
\be \label{eq:EC}
I[e_\m{}^a, \t_\m, \omega_\m{}^{ab}, \omega_\m{}^a] = \frac{\e^4}{16\pi G_N}\int dt\,d^D x\ \cE \left[ e^\m{}_a\,e^\n{}_b\,R_{\m\n}{}^{ab} - 2\,e^\m{}_a\,n^\n\,R_{\m\n}{}^a - 2\,\Lambda \right] ,
\ee
with
\begin{subequations}
\begin{align}
R_{\m\n}{}^{ab} &= 2\left(\pr_{[\m}\,\omega_{\n]}{}^{ab} + \omega_{[\m}{}^{ac}\,\omega_{\n]}{}_c{}^b + \e^2\,\omega_{[\m}{}^a\,\omega_{\n]}{}^b \right) , \\
R_{\m\n}{}^a &= 2\left(\pr_{[\m}\,\omega_{\n]}{}^a + \omega_{[\m}{}^{ab}\,\omega_{\n]}{}_b \right) .
\end{align}
\end{subequations}

The action of Eq.~\eqref{eq:EC} is invariant under local Lorentz transformations given by 
\begin{subequations}
\begin{align}
\d e_\m{}^a &= - e_\m{}^b\,\lambda^a{}_b - \e^2\t_\m\,\r^a \,,\\
\d \t_\m &= - e_\m{}^a\,\r_a \,,\\
\d\omega_\m{}^{ab} &= \pr_\m\lambda^{ab} + 2\,\omega_\m{}^{c[a}\,\lambda^{b]}{}_c + 2\,\e^2\,\omega_\m{}^{[a}\,\r^{b]} \,,\\
\d \omega_\m{}^a &= \pr_\m \r^a + \omega_\m{}^{ab}\,\r_b - \omega_\m{}^b\,\lambda^a{}_b \,,
\end{align}
\end{subequations}
where $\lambda^a{}_b$ denotes the parameter of local spatial rotations, while $\rho^{a}$ corresponds to the parameter of local Lorentz boosts.

Let us perform the following redefinitions 
\be\label{rescalingepsilon}
\omega_\m{}^a = \e^{-2}\,v_\m{}^a \,,\quad G_N = \e^2\,G_E \,,\quad \Lambda = \e^{-2}\,\Lambda_E \,.
\ee
In terms of these new quantities, the action becomes
\be\label{Iresc}
\begin{split}
I[e_\m{}^a, \t_\m, \omega_\m{}^{ab}, v_\m{}^a]& \\
&\kern-8em = \frac{1}{8\pi G_E}\int dt\,d^D x\ \cE \left[ e^\m{}_a\,e^\n{}_b\left(v_{[\m}{}^a\,v_{\n]}{}^b + \e^2\,\pr_{[\m}\,\omega_{\n]}{}^{ab} + \e^2\,\omega_{[\m}{}^{ac}\,\omega_{\n]}{}_c{}^b \right) \right. \\
&\qquad\qquad\qquad\qquad \left. - 2\,e^\m{}_a\,n^\n \left(\pr_{[\m}\,v_{\n]}{}^a + \omega_{[\m}{}^{ab}\,v_{\n]}{}_b \right) - \Lambda_E \right] ,
\end{split}
\ee
and the gauge transformations take the form
\begin{subequations}
\label{eq:transf_EC}
\begin{align}
\d e_\m{}^a &= - e_\m{}^b\,\lambda^a{}_b - \e^2\t_\m\,\r^a \,,\\
\d \t_\m &= - e_\m{}^a\,\r_a \,,\\
\d\omega_\m{}^{ab} &= \pr_\m\lambda^{ab} + 2\,\omega_\m{}^{c[a}\,\lambda^{b]}{}_c + 2\,v_\m{}^{[a}\,\r^{b]} \,,\\
\d v_\m{}^a &= \e^2\,\pr_\m \r^a + \e^2\,\omega_\m{}^{ab}\,\r_b - v_\m{}^b\,\lambda^a{}_b \,.
\end{align}
\end{subequations}
The limit $\e \rightarrow 0$ is well-defined and gives precisely the action~\eqref{eq:electric} for electric Carrollian gravity in first order form, together with the gauge transformations~\eqref{eq:transf_fields}.

A key aspect is that the gauge parameters $\lambda^a{}_b$  and $\r^a$ were not rescaled. A direct consequence is that this limit can only be applied at the level of the action and its associated gauge transformations, but not at the level of the algebra that defines the gauging. Indeed, if we apply the rescaling given in Eq.~\eqref{rescalingepsilon} to the gauge potential used in the gauging of the Poincaré algebra
\be\label{eq:PoincareConnection}
A_\m = \e \,\t_\m P_0 + e_\m{}^a P_a + \e^{-1} v_\m{}^a J_{0a} + \frac12 \omega_\m{}^{ab} J_{ab} \,,
\ee
and we require that it remains finite (and non-zero along $J_{0a}$) in the limit $\e \rightarrow 0\,$, then the boost generators of the Poincar\'e algebra must also be rescaled, as $J_{0a} = \e\,G_a\,$. Therefore, the new commutator of between two boosts reads
\be
[G_a, G_b] = \e^{-2} J_{ab} \,,
\ee
which is ill-defined in the limit when $\e \rightarrow 0\,$. This result provides an explanation for the obstacles encountered when attempting to derive electric Carrollian gravity through the gauging of the Carroll algebra~\cite{Figueroa-OFarrill:2022mcy}. However, as we will mention again in the conclusions, it is not excluded that a bigger algebra than the Carroll one, for instance based on the semi-group expansion method used in \cite{Gomis:2019nih}, provides a good starting point for deriving the action \eqref{eq:electric} together with the gauge transformations \eqref{eq:transf_fields}. 

Despite the obstructions encountered when trying to derive an action for electric Carroll gravity from a gauging procedure, some interesting connections between the first-order action \eqref{eq:electric} and the gauging of Carroll symmetry can be made.
In Appendix \ref{sec:Carrollian-vielbein-postulate}, we relate the torsion constraint arising from variation with respect to $v_\mu{}^a$ to the most general vielbein postulate compatible with the gauging of the Carroll algebra, and show that Eqs.~\eqref{eq:torsion-2} are indeed compatible with the vielbein postulate written in \cite{Hartong:2015xda}, where some part of the torsion sourced by the extrinsic curvature tensor. However, a counterpart of the vielbein postulate for the clock form is missing. This is explained in Appendix~\ref{sec:relativistic-vielbein-postulate} from the point of view of a Carrollian limit of the relativistic vielbein postulate. Moreover, it is worth noticing that the electric theory can be understood from a gauging of the Carroll algebra when supplemented with a transverse St\"uckelberg vector field $M^\mu= e^\mu{}_a\,M_a\,$, transforming inhomogeneously under Carroll boosts, as originally introduced in \cite{Hartong:2015xda} and later considered in \cite{Armas:2023dcz}. Indeed, in Appendix~\ref{sec:stueckelberg}, we propose a way to relate the fields $\omega_\mu{}^{ab}$ and $v_\mu{}^a$ to the components of a connection gauging rotations and Carroll boosts in the standard way respectively.

%%%%%%%%%%%%%%%%%%%%%%%%%%%%%%%%%%%%%%%%%%%%%%%%%%%%%%%%%%%%%%
\section{Conclusions} \label{sec:conclusions}
%%%%%%%%%%%%%%%%%%%%%%%%%%%%%%%%%%%%%%%%%%%%%%%%%%%%%%%%%%%%%%

In this article, we have presented a way to derive the electric Carrollian (or strong coupling) limit of General Relativity from a Cartan-like first-order action that is gauge-invariant under local homogeneous Carroll transformations. The key point was to consider a different realisation of Carroll boost symmetry on the auxiliary variables (the `spin connections') as compared to the usual case where one gauges the Carroll algebra along the lines of \cite{Bergshoeff:2017btm,Campoleoni:2022ebj}. 

Our result brings a positive answer to an important question, which was left open in the literature and is related to the existence of such a theory. Although an action principle in $2+1$ dimensions starting from the gauging of Carroll symmetry was shown to contain electric Carrollian gravity \cite{Hartong:2015xda}, more recent results suggested that a Cartan-like action of electric Carrollian gravity in dimensions greater than four, if it existed, could not be obtained from the gauging of the Carroll algebra \cite{Figueroa-OFarrill:2023vbj}. Our result provides a clear resolution to this tension and opens the door to constructing, for example, new theories of Carroll fermions coupled to gravity, beyond the ones presented in \cite{Bergshoeff:2023vfd}. These will be useful in order to study, e.g., the behaviour of fermionic particles near spacetime singularities. 

On the other hand, in Ref. \cite{Musaeus:2023oyp} a Palatini-type formulation for electric Carroll gravity was found from the Carrollian expansion of the Palatini action. An interesting future direction is the understanding of the relation between these two formulations when the gauge field in $v^a_\mu$ in the action \eqref{eq:electric} is considered as an independent field.

Although the realisation of Carroll boosts in the gauge transformations \eqref{eq:transf_fields} is not the standard one, it is not yet excluded that gauging a bigger algebra than Carroll would reproduce them and directly yield the first-order action \eqref{eq:electric}, upon elimination of additional auxiliary fields. Moreover, in Appendix \ref{sec:stueckelberg}, we show how to recover our first-order action and the gauge transformations \eqref{eq:transf_fields} from gauging the Carroll algebra in the presence of an extra vector field transforming under Carroll boosts as a St\"uckelberg field, making contact with previous works on Carroll geometry and field theory, see e.g., \cite{Hartong:2015xda}. Thus, the issue with the non-standard gauge transformations is most likely related to the interpretation of $M^\mu$ as being (or not) a gauge field.

Finally, let us remark that the electric action \eqref{eq:electric} can be derived by directly taking the limit $c \to 0$ of the first-order formulation of General Relativity obtained by gauging the Poincar\'e algebra. It is worth noticing that such rescaling of the fields are such that the $c \to \infty$ limit yields the \emph{Galilean} (magnetic) theory of \cite{Bergshoeff:2017btm}. Furthermore, defining $G_G=G_E/\epsilon^2$ allows to rewrite the action \eqref{Iresc} as (we take $\Lambda_E = 0$ for this discussion)
\be
I[e_\m{}^a, \t_\m, \omega_\m{}^{ab}, v_\m{}^a] = I_{Gal}[e_\m{}^a,\tau_\mu, \omega_\m{}^{ab}]
+
I_E[e_\m{}^a, \t_\m, \omega_\m{}^{ab}, v_\m{}^a] \,,
\ee
where the first term $I_{Gal}$ is the the Galilean gravity action given in Eq. (4.24) of Ref. \cite{Bergshoeff:2017btm} (with the substitution $\omega_\mu{}^{ab}\rightarrow -\omega_\mu{}^{ab}$), and the second term $I_E$ is the first-order electric Carroll gravity action~\eqref{eq:electric}. This means, in particular, that our proposed first-order electric Carroll gravity theory appears as a subleading term in the Einstein-Cartan action when the Galilean rescaling of the Poincar\'e generators and Newton's constant implied by \eqref{eq:PoincareConnection}, $P_0=\e^{-1}H\,$, $J_{0a}=\e\, G_a\,$, and $G_N=\e^4\,G_G\,$, are adopted. This is in correspondence with what happens when writing Einstein-Hilbert action in terms of pre-non-relativistic variables \cite{Hansen:2020pqs}, where the Lagrangian associated to the covariant second order form of the electric action \eqref{eq:IE} appears as a subleading term in powers of $1/c^2$ \cite{Musaeus:2023oyp} (this corresponds to Eq. (3.10) in that reference). Although still mysterious at this stage, this observation may be the starting point of a new type of electric-magnetic type of duality relating Carrollian and Galilean theories and we plan to revisit this issue in the future.

%%%%%%%%%%%%%%%%%%%%%%%%%%%%%%%%%%%%%%%%%%%
\section*{Acknowledgements}
%%%%%%%%%%%%%%%%%%%%%%%%%%%%%%%%%%%%%%%%%%%

The authors would like to thank J.~Armas, J.~Figueroa-O'Farrill, J.~Gomis, D.~Grumiller, J.~Hartong, G.~Oling, and S.~Prohazka for enlightening discussions. We especially acknowledge A.~Campoleoni and M.~Henneaux for actively taking part in the initial stages of this work and for carefully reading the manuscript. The work of S.P. was supported partly by the Fonds de la Recherche Scientifique -- FNRS under Grant No. FC.36447 and the \textit{Fonds Friedmann} run by the \textit{Fondation de l'\'Ecole polytechnique}. S.P. also acknowledges the support of the SofinaBo\"el Fund for Education and Talent. The research of A.P. is partially supported by Fondecyt grants No 1211226, 1220910 and 1230853. P.S.-R. was supported by the Austrian Science Fund (FWF), projects P 33789 and P 36619, and by the Norwegian Financial Mechanism 2014-2021 via the Narodowe Centrum Nauki (NCN) POLS grant 2020/37/K/ST3/03390. This research was partially completed during the thematic programme `Carrollian Physics and Holography' which took place in April 2024 at the Erwin Schr\"odinger Institute of Vienna, where a preliminary version of these results was also presented. S.P. and P.S.-R. thank the University of Edinburgh for hospitality during the development of this paper. P.S.-R. acknowledges the Dublin Institute for Advanced Studies (DIAS) for support during the final stage of this work.

\appendix

%%%%%%%%%%%%%%%%%%%%%%%%%%%%%%%%%%%%%%%%%%%
\section{Relation with the gauging of the Carroll algebra} \label{sec:gauging}
%%%%%%%%%%%%%%%%%%%%%%%%%%%%%%%%%%%%%%%%%%%

%%%%%%%%%%%%%%%%%%%%%%%%%%%%%%%%%%%%%%%%%%%
\subsection{Most general Carrollian vielbein postulate} \label{sec:Carrollian-vielbein-postulate}

The equation of motion Eq.~\eqref{eq:torsion-2} is only a projection of the condition of vanishing torsion obtained by gauging the Carroll algebra \cite{Bergshoeff:2017btm,Campoleoni:2022ebj}. In the usual setup, the term $v_{[\m}{}^a \t_{\n]}$ can be thought of as a source of torsion \cite{Hartong:2015xda}.

More precisely, upon imposing Eq.~\eqref{vandomega}, we find that Eqs.~\eqref{eq:torsion-2} read
\begin{subequations} \label{eq:vielbein-postulates}
\begin{align}
n^\m \left( \pr_{[\m} e_{\n]}{}^a + \omega_{[\m}{}^{ab} e_{\n]}{}_b + \t_{[\m} e_{\n]}{}_b K^{ab} \right) &\approx 0 \,,\\
e^\m{}_a \left( \pr_{[\m} e_{\n]}{}^a + \omega_{[\m}{}^{ab} e_{\n]}{}_b + \t_{[\m} e_{\n]}{}_b K^{ab} \right) &\approx 0 \,.
\end{align}
\end{subequations}
In order to compare these equations with the vielbein postulate, we need to get rid of the projection, which is done by considering the most general Ansatz compatible with Eqs.~\eqref{eq:vielbein-postulates}
\be \label{eq:unprojected-vielbein-postulate}
\pr_{[\m} e_{\n]}{}^a + \omega_{[\m}{}^{ab} e_{\n]}{}_b + \t_{[\m} e_{\n]}{}_b K^{ab} \approx e_{[\m}{}^b e_{\n]}{}^c A^a{}_{bc} \,,
\ee
for an arbitrary tensor $A^a{}_{bc}\,$, antisymmetric in its last two indices, traceless $A^a{}_{ba} = 0\,$, and transforming as a tensor under local rotations only
\be
\d A^a{}_{bc} = \lambda^a{}_d A^d{}_{bc} + \lambda_b{}^d A^a{}_{dc} + \lambda_c{}^d A^a{}_{bd} \,.
\ee
This tensor $A^a{}_{bc}$ carries $\frac12 (D-2)D(D+1)$ components, which corresponds to the components of $\omega_\m{}^{ab}$ that are left free due to the projections \eqref{eq:vielbein-postulates}.

Comparing Eq.~\eqref{eq:unprojected-vielbein-postulate} with the vielbein postulate\footnote{There is a slight change of convention: $(\omega_\m{}^{ab})_\text{here} = - (\Omega_\m{}^{ab})_\text{there}\,$, $(v_\m{}^a)_\text{here} = - (\Omega_\m{}^a)_\text{there}$ and $(n^\mu)_\text{here} = - (v^\mu)_\text{there}\,$, so that $(K_{\m\n})_\text{here} = -(K_{\m\n})_\text{there}\,$.} (2.22) of \cite{Hartong:2015xda}
\be
\pr_\m e_\n{}^a + \omega_\m{}^{ab}e_\n{}_b - \G_{\m\n}^\r e_\r{}^a = 0 \,,
\ee
we arrive at the identification
\be \label{eq:identification}
\Gamma_{[\m\n]}^\r g_{\r\s} = - \t_{[\m} K_{\n]\sigma} + X^\r_{[\m\n]} g_{\r\s} \,,
\ee
where we defined $X^\r_{[\m\n]} = e^\r{}_a e_\m{}^b e_\n{}^c A^a{}_{[bc]}\,$.

Eq.~\eqref{eq:identification} is compatible with Eqs.~(2.36) and (2.38) of \cite{Hartong:2015xda}. Note however that our $X^\r_{[\m\n]}$ is not the most general arbitrary tensor allowed by the gauging of the Carroll algebra, it is $\frac12 D(D+1)$ components short of the most general one described in \cite{Hartong:2015xda}. These extra components correspond to the antisymmetric projection of $h_{\s\r}\,n^\n\,X^{\r}_{[\m\n]}$ allowed by Eq.~(2.37) of \cite{Hartong:2015xda} (which are automatically zero here since $n^\n\,X^\r_{[\m\n]} = 0$) accounting for $\frac12 D(D-1)$ components, and to the trace $X^\r_{[\r\m]}\,$, which is a purely spatial vector accounting for $D$ components.

The comparison with \cite{Hartong:2015xda} stops here, since a vielbein postulate (or even a torsion constraint) for $\t_\m$ is missing in the list of equations of motion obtained by varying our action, so that the temporal projection of the connection $\t_\r\,\G^\r_{\m\n}$ is left completely free. Note, however, that if one considered Eq.~(2.43) of \cite{Hartong:2015xda} and assumed that our $v_\m{}^a$ is the spin connection associated to Carrollian boosts, then $\t_\r\,\G^\r_{\m\n}$ should be fully determined in terms of $\pr_\m \t_\n$ and $K_{\m\n}\,$. This apparent tension is resolved by the fact that $v_\m{}^a$ is in fact not the connection gauging the generator of Carroll boosts, as explained in sections~\ref{sec:introduction} and \ref{sec:Electric_action_from_a_limit}, which is manifest in the transformation law given in Eq.~\eqref{eq:deltav}.

%%%%%%%%%%%%%%%%%%%%%%%%%%%%%%%%%%%%%%%%%%%
\subsection{From a limit of the relativistic vielbein postulate} \label{sec:relativistic-vielbein-postulate}

Since our first-order action can be obtained from a limit of the Einstein-Cartan action after a suitable rescaling, one can wonder what happens to the vielbein postulate in this limit.

Let us start from the relativistic setup with metric and co-metric given by
\be
G_{\m\n} = g_{\m\n} - \e^2\,\t_\m\,\t_\n \,,\quad G^{\m\n} = g^{\m\n} - \e^{-2}\,n^\m\,n^\n \,,
\ee
where $g_{\m\n} = e_\m{}^a\,e_\n{}^b\,\d_{ab}\,$. The associated Levi-Civita connection reads
\be
\G_{\m\n}^\r = \frac12\,G^{\r\s}\left(\pr_\m\,G_{\n\s} + \pr_\n\,G_{\m\s} - \pr_\s\,G_{\m\n} \right) ,
\ee
where, like in section~\ref{sec:Electric_action_from_a_limit}, we introduced the speed of light as $\e\,$. The relativistic `vielbein postulate',\footnote{In the relativistic case, where the Christoffel symbols are given by the Levi-Civita connection, it is not a postulate but an identity.} split along $e_\m{}^a$ and $\t_\m\,$, reads
\begin{subequations} \label{eq:vielbein-postulates-relativistic}
\begin{align}
\pr_\m\,e_\n{}^a + \omega_\m{}^{ab}\,e_\n{}_b + \e^2\,\omega_\m{}^a\,\t_\n - \G_{\m\n}^\r\,e_\r{}^a &= 0 \,,\\
\pr_\m\,\t_\n{} + \omega_\m{}^a\,e_\n{}_a - \G_{\m\n}^\r\,\t_\r &= 0 \,.
\end{align}
\end{subequations}
Expanding the Levi-Civita connection in powers of $\epsilon$, we obtain
\be
\Gamma^\rho_{\mu\nu} = - \epsilon^{-2} n^\rho K_{\mu\nu} + \gamma^\rho_{\mu\nu} + n^\rho \left(\partial_{(\mu} \tau_{\nu)} - \tau_{(\mu} \mathcal L_n \tau_{\nu)}\right) + \mathcal O(\epsilon^2) \,,
\ee
where we defined $\g_{\m\n}^\r = \frac12\,g^{\r\s} \left( \pr_\m\,g_{\n\s} + \pr_\n\,g_{\m\s} - \pr_\s\,g_{\m\n} \right)$.\footnote{Taking the limit $\e \rightarrow 0$ and considering the leading order part, the vielbein postulates \eqref{eq:vielbein-postulates-relativistic} become \mbox{$\pr_\m\,e_\n{}^a + \omega_\m{}^{ab}\,e_\n{}_b - \g_{\m\n}^\r\,e_\r{}^a = 0$} and $K_{\mu\nu} = 0\,$. In the first equation we used the fact that \mbox{$\Gamma^\rho_{\mu\nu}\,e_\rho{}^a = \gamma^\rho_{\mu\nu}\, e_\rho{}^a + \mathcal O(\epsilon^2)$} and in the second that $\Gamma^\rho_{\mu\nu}\,\tau_\rho = - \epsilon^{-2} K_{\mu\nu} + \mathcal O(1)\,$. In the limit, both equations are compatible with the magnetic theory defined in \cite{Campoleoni:2022ebj}.}

Performing the redefinition $\omega_\m{}^a = \e^{-2}\,v_\m{}^a$ identified in Eq.~\eqref{rescalingepsilon}, the limit $\e \rightarrow 0$ reads
\begin{subequations}
\begin{align}
\pr_\m\,e_\n{}^a + \omega_\m{}^{ab}\,e_\n{}_b + v_\m{}^a\,\t_\n - \g_{\m\n}^\r\,e_\r{}^a &= 0 \,,\label{VP1}\\
v_\m{}^a\,e_\n{}_a + K_{\mu\nu} &= 0 \,.
\end{align}
\end{subequations}
The latter equation is equivalent to~\eqref{vandomega}, while the first one is again a vielbein postulate with a source of torsion given by the third term of the left-hand side. Eq.~\eqref{VP1} is compatible with Eqs.~\eqref{eq:unprojected-vielbein-postulate} and \eqref{eq:identification}, provided that any ambiguity encoded in the arbitrary piece $X^\r_{[\m\n]}$ vanishes.

All in all, we find that the torsion constraints arising in the first-order formulation of the electric theory are compatible with the vielbein postulates defined in \cite{Hartong:2015xda}, but is not equivalent to its most general form since it does not contain all allowed arbitrary pieces of torsion.

This can be explained due to the origin of the electric theory as a certain limit of relativistic Einstein-Cartan theory, where the affine connection is completely determined in terms of metric quantities. The ambiguity related to the projection in Eqs.~\eqref{eq:vielbein-postulates} and to the fact that the field $\t_\m$ does not satisfy any form of torsion constraint still persist in this limit, but do not enter the action \eqref{eq:electric}, which reduces to the action for electric Carrollian gravity upon switching to metric variables.

%%%%%%%%%%%%%%%%%%%%%%%%%%%%%%%%%%%%%%%%%%%
\subsection{Using a St\"uckelberg field} \label{sec:stueckelberg}

We can introduce a St\"uckelberg field $M^a$ and a new set of spin connections $\tilde v_\mu{}^a$, $\tilde \omega_\mu{}^{ab}$ transforming under Carroll boosts like the spin connections that follow from gauging the Carroll algebra
\be
    \delta \tilde v_\mu{}^a=\tilde D_\mu \rho^a - \lambda^a{}_b\, \tilde v_\mu{}^b ,\qquad \delta \tilde \omega_\mu{}^{ab}=\tilde D_\mu \lambda^{ab},\qquad\delta M^a= -\lambda^a{}_b\,M^b + \rho^a \,,
\ee
where $\tilde D$ is the derivative acting covariantly on frame indices using the spin connection $\tilde \omega_\mu{}^{ab}$, e.g., $\tilde D_\mu \rho^a = \partial_\mu \rho^a + \tilde\omega_\mu{}^a{}_b\,\rho^b$. Let us define the new fields\footnote{The introduction of a vector field $M^a$ transforming as a shift under Carroll boosts has been considered previously in section 2.3 of \cite{Hartong:2015xda}. Moreover, the introduction of the term $\tilde D_\mu M^a$ in the spin connection related to Carroll boosts has been motivated before when the extrinsic curvature of the Carroll manifold does not vanish, see, e.g., Eq.~(3.47) of \cite{Bergshoeff:2017btm}.}
\be
    v_\mu{}^a:=\tilde v_\mu{}^a-\tilde D_\mu M^a,
    \qquad
    \omega_\mu{}^{ab}:=\tilde \omega_\mu{}^{ab}+2\,v_\mu{}^{[a}M^{b]}=\tilde \omega_\mu{} ^{ab}+2\,\tilde{v}_\mu{}^{[a}M^{b]}-2\,\tilde D_\mu  M^{[a} M^{b]} \,.
\ee
One can check that, under homogeneous Carroll transformations, $v^a$ and $\omega^{ab}$ transform as the spin connections in Eq.  \eqref{eq:transf_fields}, i.e.
\be
    \delta v_\mu{}^a= - \lambda^a{}_b v_\mu{}^b \,,\qquad
    \delta \omega_\mu{}^{ab}=D_\mu \lambda^{ab} + 2v_\mu{}^{[a} \rho^{b]} \,,
\ee
where $D_\mu$ is the frame covariant derivative built using the spin connection $\omega_\mu{}^{ab}$.
The antisymmetrized covariant derivative of $v_\mu{}^a$ w.r.t. $\omega_\mu{}^{ab}$ can be written in terms of the tilded fields as
\be
\begin{split}
    D_{[\mu} v_{\nu]}{}^a&=\tilde D_{[\mu} v_{\nu]}{}^a+2\left(\tilde v_{[\mu}{}^{[a}-\tilde D_{[\mu}  M^{[a}\right) M^{b]} v_{\nu] b} \\
    &
    =\tilde D_{[\mu} \tilde v_{\nu]}{}^a-\frac12\tilde R_{\mu\nu}{}{}^a{}_b M^b-2
    \left(\tilde v_{[\mu b}
    -\tilde D_{[\mu} M_b\right)
  \left(\tilde v_{\nu]}^{[a}
    -\tilde D_{\nu]} M^{[a}\right)
    M^{b]} \,,
\end{split}
\ee
where $\tilde R_{\mu\nu}{}{}^{ab}=2\,\partial_{[\mu}\tilde \omega_{\nu]}{}^{ab}+2\,\tilde \omega_{[\mu}{}^a{}_c \,\tilde\omega_{\nu]}{}^{cb}$.
Therefore, when written in terms of the tilded spin connections, the action \eqref{eq:electric} takes the form
\be \label{eq:electricStuckelberg}
\begin{split}
    &I_E[e_\m{}^a, \t_\m, \tilde\omega_\m{}^{ab}, \tilde v_\m{}^a,  M^a] \\
    &= \frac{1}{8\pi G_E}\int dt\,d^D x\ \cE \left[ e^\m{}_a\,e^\n{}_b\,\tilde v_{[\m}{}^a\,\tilde v_{\n]}{}^b - 2\,e^\m{}_a\,n^\n\left(\pr_{[\m}\, \tilde v_{\n]}{}^a + \tilde\omega_{[\m}{}^{ab}\, \tilde v_{\n]}{}_b\right) - \Lambda_E \right] \\
    &\quad + I_\text{St\"uckelberg} \,,
\end{split}
\ee
where the action $I_\text{St\"uckelberg}$ is given by
\be \label{eq:actionforpsi}
\begin{split}
    I_\text{St\"uckelberg}
    &= \frac{1}{8\pi G_E}\int dt\,d^D x\ \cE \bigg[  
    e^\m{}_a\,e^\n{}_b\left(\tilde D_{[\mu} M^a \,\tilde D_{\nu]} M^b-2\tilde D_{[\mu} M^a\,\tilde v_{\nu]}{}^b 
    \right) \\
    &\quad-\,e^\m{}_a\,n^\n \,\tilde R_{\mu\nu}{}{}^{ab} M_b+\,4\,e^{\m a}\,n^\n \left(\tilde v_{[\mu}{}^b-\tilde D_{[\mu} M^b\right)\left(\tilde v_{\nu][a}-\tilde D_{\nu]} M_{[a}\right) M_{b]}\bigg] .
\end{split}
\ee
We can construct this action from a gauging procedure using the Carrollian gauge connection
\be
\begin{split}
    A_\mu&= \tau_\mu H + e_\mu{}^a P_a + \tilde v_\mu{}^a C_a +\frac12 \tilde \omega_\mu{}^{ab} J_{ab} \\
    &= \tau_\mu H + e_\mu{}^a P_a + \left(v_\mu{}^a + D_\mu M^a-2\, v_\mu{}^{[a}M^{b]}M_b\right) C_a +\frac12 \left(\omega_\mu{}^{ab} -2\, v_\mu{}^{[a} M^{b]}\right)J_{ab} \,.
\end{split}
\ee
Note that if introduce a vector $M^\mu$ transforming as in \cite{Hartong:2015xda}
\be
\delta M^\mu = e^\mu{}_a \,\rho^a \,,
\ee
we can define
\be
M^a=M^\mu \,e_\mu{}^a \,.
\ee
However, unlike the extra field introduced in \cite{Hartong:2015xda}, $M^\mu$ can be taken to be purely transverse from the beginning $M^\mu \,\tau_\mu=0\,$, which resembles the Goldstone field $\theta^\mu$ considered in \cite{Armas:2023dcz} upon absorbing the time-like component using the extra symmetry $\delta \theta^\mu = \chi n^\mu\,$.

%%%%%%%%%%%%%%%%%%%%%%%%%%%%%%%%%%%%%%%%%%%
\section{Alternative covariant first-order form of the electric action}
\label{AppFirstOrder}
%%%%%%%%%%%%%%%%%%%%%%%%%%%%%%%%%%%%%%%%%%%

Varying the action \eqref{eq:electric} with respect to $M^{ab}{}_a\,$, $N^{ab}\,$, $S_{[ab]}$ and $U_b$ yields the closed set of equations
\be \label{eq:torsion-incomplete}
U_b \approx 0 \,,\quad S_{[ab]} \approx 0 \,,\quad N^{ab} \approx 2\,n^\m\,e^\n{}^{[a}\,\pr_{[\m}\,e_{\n]}{}^{b]} \,,\quad M^{ab}{}_a \approx 2\,e^\m{}^b\,e^\n{}_a\,\pr_{[\m}\,e_{\n]}{}^a \,.
\ee
By replacing them back into the original action, we recover a hybrid form of the electric action, which is first-order in time derivatives, but formulated in terms of the metric $g_{\m\n}$ and an auxiliary field $S_{\m\n} = v_{(\m}{}^a\,e_{\n)}{}_a$
\be
I_E[g_{\m\n}, n^\m, S_{\m\n}] \approx \frac{1}{8\pi G_E}\int dt\,d^D x\ \cE \left[ S_{[\m}{}^\m S_{\n]}{}^\n + 2\,K_{[\m}{}^\n\,S_{\n]}{}^\m -\,\Lambda_E \right] ,
\ee
where curved indices are lowered and raised using the degenerate metric $g_{\m\n}$ or its pseudo-inverse $g^{\m\n}\,$, which does not introduce any projection because both $S_{\m\n}$ and $K_{\m\n}$ are transverse to the time-like component of the vielbein $n^\m\,$. Defining
\be
P^{\m\n} = S^{\m\n} - g^{\m\n} S_\a{}^\a \,,
\ee
we see that $P^{\m\n}$ indeed plays the role of `canonical momenta' associated to the metric, in the sense that
\be \label{eq:electric-hybrid}
I_E[g_{\m\n}, n^\m, P^{\m\n}] \approx \frac{1}{8 \pi G_E} \int dt\,d^D x\ \cE \left[ K_{\m\n} \, P^{\m\n} - G_{\m\n\r\s}\,P^{\m\n}\,P^{\r\s} - \Lambda_E \right] ,
\ee
with $G_{\m\n\r\s} = \frac12 \left[ g_{\m\r} g_{\n\s} - \frac{1}{D-1} g_{\m\n} g_{\r\s} \right]$. By eliminating $P^{\m\n}$ through its own equation of motion, we recover the electric action \eqref{eq:El_action}. This action is manifestly invariant under local rotations, and was proposed in \cite{Campoleoni:2022ebj} as a covariantisation of the Hamiltonian formulation of electric Carroll gravity. Note that, in this form, the gauge transformation associated to Carroll boosts acts in a non-differential way
\be
\d_\rho S^{\m\n} = n^{(\m} \r^{\n)} \,,\quad \d_\rho P^{\m\n} = n^{(\m} \tilde\r^{\n)} \,,
\ee
where we posed
\be
\r^\m = 2\,\r_a\,e^\s{}_b\,e^\m{}^b\,v_\s{}^a\,,\quad \tilde \r^\m = 4\,\r_a\,e^\s{}_b\,e^\m{}^{[b}\,v_\s{}^{a]} \,,
\ee
verifying $\t_\m\,\r^\m = 0$ and $\t_\m\,\tilde\r^\m = 0\,$. The existence of a gauge symmetry of the electric action in the form of Eq.~\eqref{eq:electric-hybrid} was already observed in \cite{Campoleoni:2022ebj}, although its origin as a gauge symmetry in the first order formulation was not manifest.

%%%%%%%%%%%%%%%%%%%%%%%%%%%%%%%%%%%%%%%%%%%
\section{Equivalence to the electric theory in Hamiltonian form}
\label{AppHamiltonian}
%%%%%%%%%%%%%%%%%%%%%%%%%%%%%%%%%%%%%%%%%%%

We can directly recover the Hamiltonian formulation of electric Carroll gravity by starting from the action~\eqref{eq:electric} and choosing the so-called time gauge, i.e.
\be\label{ParNRfields}
\bal
&\tau_\mu=(N,0)\,,\quad n^\mu=(N^{-1},-N^i N^{-1}) \,,
\\[5pt]
& e_\mu{}^a =(N^j e_j{}^a, e_i{}^a)\,,\quad e^\mu{}_a =(0, e^i{}_a) \,.
\eal
\ee
where $N$ and $N_i$ are identified as the lapse and shift functions of the ADM decomposition. This gauge is reached by making use of the gauge variation of $e_\m{}^a$ and $\t_\m$ with parameter $\rho^a$, which takes the same form as in the magnetic case.
In terms of these new variables, the determinant $\mathcal E$ reduces to
\be
\mathcal E = \frac{N}{D!}\,\ve^{0\,i_1 \,\cdots\, i_D} \, \ve_{a_1\,\cdots\, a_D}\, e_{i_1}{}^{a_1} \,\cdots\, e_{i_D}{}^{a_D} = N\, e \,.
\ee
In the time gauge, the action~\eqref{eq:electric} takes the form
\be\label{MMactionExt3}
\begin{split}
&I_E[e_i{}^a, N^i, N, v_\m{}^a, \omega_\m{}^{ab}] \\
&\ = \frac{1}{8\pi G_E} \int dt\,d^D x\ e \left[ N\,e^i{}_a\,e^j{}_b\,v_{[i}{}^a\,v_{j]}{}^b + 2\,e^i{}_a\,D_{[0}\,v_{i]}{}^a - 2\,N^i\,e^j{}_a\,D_{[i}\,v_{j]}{}^a -N\,\Lambda_E \right] ,
\end{split}
\ee
where $D_\mu$ denotes the covariant derivative with respect to  $\omega_\mu{}^{ab}\,$.

Varying the action \eqref{MMactionExt3} with respect to $\omega_0{}^{ab}$ and $e^i{}_b\,\omega_i{}^{ab}\,$, gives the conditions $S^{[ab]} \approx 0$ and $v_0{}^a \approx N^i v_i{}^a$. Additionally, the variation with respect to $v_0{}^a$ gives the following equation of motion
\be
\pr_i e^i{}_a + e^i{}_a\,e^j{}_b\,\pr_i e_j{}^b + e^i{}_b\,\omega_i{}^{ab} \approx 0 \,.
\ee
In the case of magnetic Carroll gravity \cite{Campoleoni:2022ebj}, one was able to derive a vielbein postulate for the spatial, non-degenerate part of the vielbein, but here we do not obtain exactly the standard vielbein postulate $\nabla_i\,e_j{}^b - \omega_i{}^{bc}\,e_{jc} = 0$, but only a projection of it, namely $e^i{}_b\left(\nabla_i\,e_j{}^b - \omega_i{}^{bc}\,e_{jc}\right) = 0\,$. Here, $\nabla_{i}$ denotes the covariant derivative with respect to the Levi-Civita connection $\gamma_{ij}^k\,$ given by
\be \label{Levi-Civita}
\g_{ij}^k = \frac12\,g^{kl} \left( \pr_i g_{lj} + \pr_j g_{il} - \pr_l g_{ij} \right) ,
\ee
which is exclusively defined in terms of the non-degenerate spatial metric $g_{ij}=\delta_{ab}\,e_i{}^a\,e_j{}^b$ and its inverse.

Taking the variation of the action \eqref{MMactionExt3} with respect to $e^i{}^{[a}\,v_i{}^{b]}$, and using the equation of motion $S^{[ab]} \approx 0$, one finds
\be
e^i{}^{[a}\,\dot e_i{}^{b]} - \pr_i N^j e^i{}^{[a} e_j{}^{b]} + N^i \pr_i e^j{}^{[a} e_j{}^{b]} = \omega_0{}^{ab} - N^i \omega_i{}^{ab} \,.
\ee
If the above conditions are replaced in the action \eqref{MMactionExt3}, and boundary terms are neglected, one obtains precisely the action for electric Carroll gravity in Hamiltonian form 
\cite{Henneaux:1979vn,Henneaux:2021yzg}
\be\label{MMCarrollExt5}
\bal
&I_E[g_{ij}, N_i, N, \pi^{ij}] =\int dt\,d^D x \left[ \pi^{ij}\,\dot g_{ij} - N\,\mathcal H - N^i\,\mathcal H_i \right] ,\\[6pt]
&\mathcal H= \frac{16\pi G_E}{\sqrt g}\left(\pi^{ij}\,\pi_{ij}-\frac{1}{D-1}\pi^2\right)+\frac{\sqrt g \Lambda_E}{8\pi G_E} \,,\qquad \mathcal H_i = - 2\,\nabla_j\,\pi^{ij} \,,
\eal
\ee
where 
\be\label{momentum}
\pi^{ij}=\frac{e}{16\pi G_E} \left(e^i{}_b\,v^{jb} - g^{ij}e^k{}_a\,v_k{}^a\right)
=\frac{e}{16\pi G_E} \left(S^{ab}e^i{}_a e^j{}_b - g^{ij}S\right) , 
\ee
with $S=\delta_{ab} S^{ab}\,$. 

Alternatively, the form of the action \eqref{MMCarrollExt5} can be recovered by using an ADM decomposition in the action~\eqref{eq:electric-hybrid}.

%%%%%%%%%%%%%%%%%%%%%%%%%%%%%%%%%%%%%%%%%%%%%%%%%%%%%%%%

%\bibliographystyle{JHEP}

\end{document}